\documentclass[epj, nopacs]{svjour}
\usepackage{graphics}
\usepackage[
    locale=US,
    separate-uncertainty=true,
    per-mode=fraction,    
    retain-unity-mantissa = false,
]{siunitx}
\DeclareSIUnit\clight{\text{\ensuremath{c}}}
\usepackage{amsmath}
\usepackage{float}
\usepackage{scrhack}
\usepackage{hyperref}
\usepackage[
backend=biber,
style=numeric,
sorting=none,
natbib
]{biblatex}
\usepackage[T1]{fontenc}
\usepackage{microtype}
\usepackage{booktabs}
\usepackage{multirow}

\begin{document}
\title{Simulation of Deflection Uncertainties on Directional Reconstructions of Muons Using PROPOSAL}
% \subtitle{Do you have a subtitle?\\ If so, write it here}
\author{Pascal Gutjahr\inst{1}\fnmsep\thanks{\email{pascal.gutjahr@udo.edu} (corresponding author)} \and Jean-Marco Alameddine\inst{1} \and Alexander Sandrock\inst{2} \and Jan Soedingrekso\inst{1} \and Mirco Hünnefeld\inst{1} \and Wolfgang Rhode\inst{1}  
}
\institute{Department of Physics, TU Dortmund University, Otto-Hahn-Straße~4, Dortmund 44227, Germany \and 
          Faculty of Mathematics and Natural Sciences, University of Wuppertal, Gaußstraße~20, Wuppertal 42119, Germany}
\date{Received: 23. August 2022 / Revised version: 10. January 2023}
\abstract{
  Large scale neutrino detectors and muography rely on the muon direction in the detector to infer the muon's or parent neutrino's origin. However, muons accumulate deflections along their propagation path prior to entering the detector, which may need to be accounted for as an additional source of uncertainty. 
  In this paper, the deflection of muons is studied with the simulation tool PROPOSAL, which accounts for multiple scattering and deflection on stochastic interactions. Deflections along individual interactions depend on the muon energy and the interaction type, and can reach up to the order of degrees -- even at TeV to PeV energies. The accumulated deflection angle can be parametrized in dependence of the final muon energy, independent of the initial muon energy. The median accumulated deflection of a propagated muon with a final energy of $\SI{500}{\giga\electronvolt}$ is $\theta_{\text{acc}} = \SI{0.10}{\degree}$ 
  with a $\SI{99}{\percent}$ central interval of $[\SI{0.01}{\degree}, \,\SI{0.39}{\degree}]$. 
  This is on the order of magnitude of the directional resolution of present neutrino detectors. Furthermore, comparisons with the simulation tools MUSIC and \textsc{Geant4} as well as two different muon deflection measurements are performed.
%
% \PACS{
%       {PACS-key}{discribing text of that key}   \and
%       {PACS-key}{discribing text of that key}
%      } % end of PACS codes
} %end of abstract
\authorrunning{P. Gutjahr et al.}
\titlerunning{Muon Deflection Simulation Using PROPOSAL}
\maketitle
\section{Introduction}\label{sec:introduction}

The directional reconstruction of muons is an essential task for muography or large scale neutrino detectors such as IceCube 
\cite{IceCube_Instrumentation} or KM3NeT \cite{KM3NeT_Design}. In both cases, the muon direction is measured at its crossing through the instrumented volume, which is then utilized to infer its origin or the origin of the parent neutrino. 
However, muons may propagate many kilometers prior to entering the detector while interacting with the surrounding medium. 
Along their propagation, muons can undergo many of thousands 
of interactions, depending on their energy and propagation distance. 
These interactions can lead to a deflection of the muon that may need to be accounted for as an additional source of uncertainty in these measurements. 
Current angular resolutions are above
$\SI{0.1}{\degree}$ for  
$\si{\tera\electronvolt}$ to $\si{\peta\electronvolt}$ energies in IceCube 
\cite{IceCube_Resolution2021} 
and below 
$\SI{0.2}{\degree}$ for energies greater than $\SI{10}{\tera\electronvolt}$ in 
KM3NeT/ARCA (part of KM3NeT dedicated to search for very high-energetic neutrinos) \cite{KM3NeT_Resolution2021}.

To study the impact of the muon deflection on the angular resolution 
of current neutrino detectors, 
the paper is structured as follows: in section~\ref{sec:proposal},
the lepton propagator PROPOSAL is briefly described. In section~\ref{sec:defl_per_int},
PROPOSAL \cite{koehne2013proposal, dunsch_2018_proposal_improvements} is used to study the muon deflection per interaction.
The accumulated deflection is analyzed and compared to the propagation codes
MUSIC \cite{MUSIC, comparison_MUSIC_GEANT4_2009} and \textsc{Geant4} \cite{GEANT4_standard, GEANT4} and data from two experiments in section~\ref{sec:accum_defl}. The findings of this study
are summarized in section~\ref{sec:conclusion}.

\section{Overview of the Simulation Tool PROPOSAL}\label{sec:proposal}

The tool PROPOSAL \cite{koehne2013proposal, dunsch_2018_proposal_improvements} propagates charged leptons and photons through media and is 
used in this paper to simulate the deflection of muons. All relevant muon interaction types 
as bremsstrahlung \cite{KKP_1995, Bremsstrahlung_KKP}, photonuclear interaction \cite{Abramowicz_1997} with 
shadowing \cite{ButkevichMikheyev_2002}, electron pair production \cite{epair_kokoulin_petrukhin} with corrections for the 
interaction with atomic electrons \cite{epair_kelner}, 
ionization described by the Bethe-Bloch formula with corrections for muons \cite{Rossi}, 
and the decay are provided by PROPOSAL. The interaction processes are sampled by their cross section.
Since energy losses
with the massless photon as secondary particle can be arbitrarily small, an energy cut is introduced to avoid an infinite number of bremsstrahlung interactions 
and furthermore to increase the runtime performance. 
The cut is applied with a minimum energy loss
\begin{equation}
    E_{\text{loss,min}} = \min{(e_{\mathrm{cut}}, E \cdot v_{\mathrm{cut}})}\,,
\end{equation}
using two parameters -- a total and a relative energy cut denoted as 
$e_{\mathrm{cut}}$ and $v_{\mathrm{cut}}$ with the energy $E$ of the particle 
directly before the interaction. 
By the introduction of 
this energy cut, the next significant energy loss with 
$E_{\mathrm{loss}} \geq E_{\text{loss,min}}$ 
is treated as a stochastic energy loss in the propagation. 
All energy losses with $E_{\mathrm{loss}} < E_{\text{loss,min}}$ between 
two stochastic losses are accumulated and lost continuously, denoted 
as continuous energy loss.
The methodical uncertainties are small 
for a relative energy cut $v_{\mathrm{cut}}\ll 1$, however, using a small energy 
cut increases the runtime.
Typically, a relative energy cut of $v_{\mathrm{cut}} \leq \num{0.05}$ 
is chosen which enables accurate propagations at low runtimes. The total energy 
cut depends on the minimum visible energy loss in the detector. 
It is often set to $e_{\mathrm{cut}} = \SI{500}{\mega\electronvolt}$.
The 
propagation process is defined by an initial muon energy $E_{\text{i}}$ and 
two stopping criteria -- a final energy $E_{\text{f,\,min}}$ and a 
maximum propagation distance $d_{\text{max}}$. If the last interaction of 
a propagation is sampled by a stochastic interaction, the true final energy 
$E_{\text{f}}$ can become lower. 
Since muons are unstable, a decay leads to a premature 
stop, which is negligible for high energies.

In PROPOSAL, the deflections for stochastic interactions are parametrized by Van Ginneken 
in Ref.~\cite{Van_Ginneken} with a direct calculation of the deflection in 
ionization using four-momentum conservation. 
Furthermore, there are parametrizations for stochastic deflections given in \textsc{Geant4} \cite{GEANT4_standard, GEANT4} 
for bremsstrahlung and photonuclear interaction, which 
are also available.
To estimate the deflection along 
a continuous energy loss, multiple scattering described by Molière 
(MSM) \cite{moliere_scattering} and the Gaussian approximation by 
Highland (MSH) \cite{HIGHLAND_1975}
can be chosen. 
MSM results as a summation of elastic scatterings of one particle 
at another particle, called single scattering. Thus, the muon is deflected by a 
single angle for each continuous loss, analogous to a stochastic loss.
The orientation of the deflection in the plane perpendicular to the muon direction is 
sampled uniformly between $0$ and $2\mathrm{\pi}$.
The latest updates with a detailed description of the whole tool can be found 
in Ref.~\cite{phd_soedingrekso}. 
The stochastic deflections have been implemented in PROPOSAL recently and they are 
described and studied in Ref.~\cite{Gutjahr_2021}.
A publication describing the 
updates in PROPOSAL is in preparation 
\cite{alameddine_et_al}.
All simulations are done with PROPOSAL $7.3.1$.

\section{Muon Deflection per Interaction}\label{sec:defl_per_int}
\begin{figure*}
    \centering 
    \includegraphics{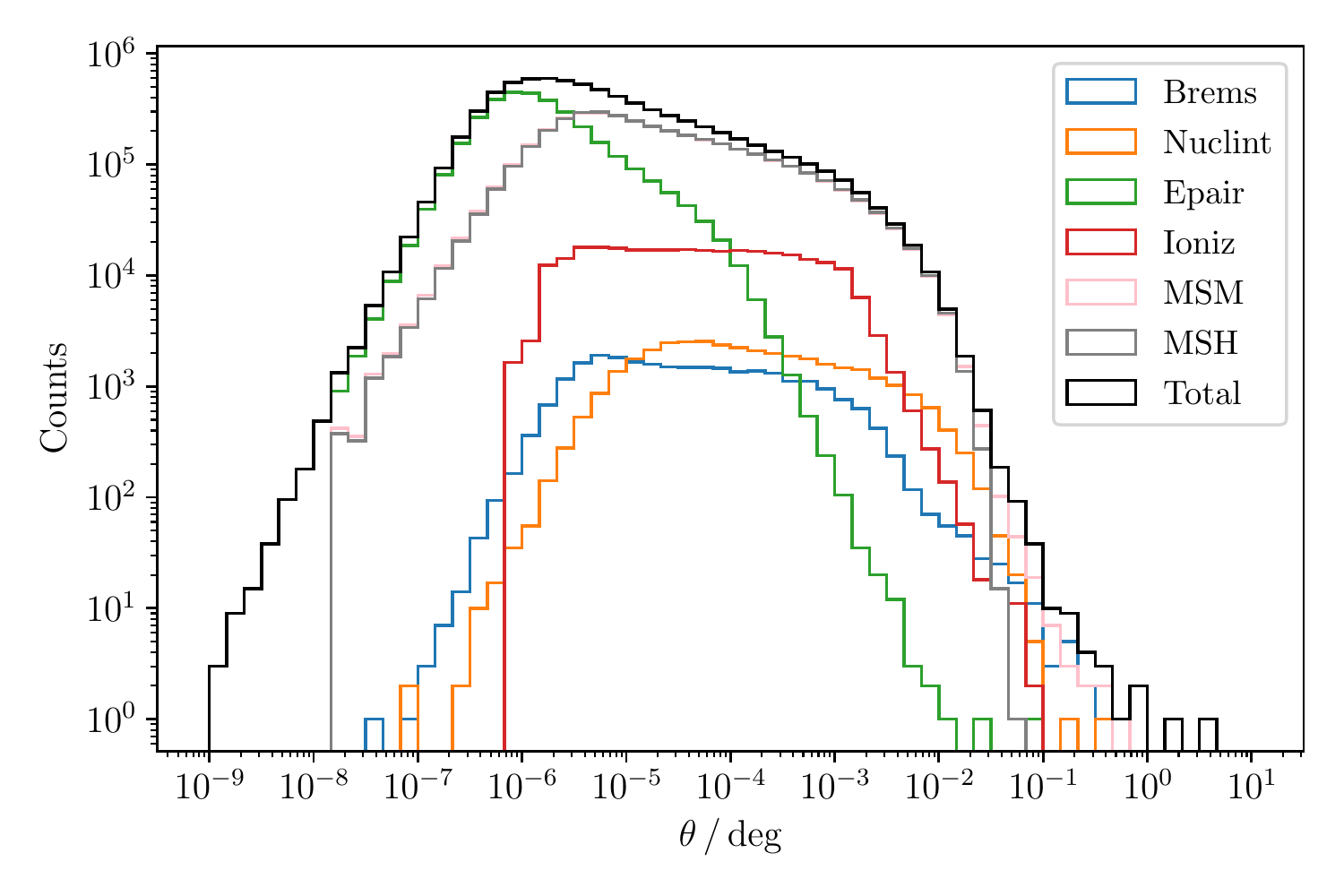}
    \caption{The muon deflection $\theta$ per interaction in degree is shown for different mechanisms. The propagation is done for $\num{1000}$ 
    muons from $E_{\text{i}} = \SI{1}{\peta\electronvolt}$ to $E_{\text{f,\,min}} = \SI{1}{\tera\electronvolt}$ using $e_{\mathrm{cut}} = \SI{500}{\mega\electronvolt}$ and $v_{\mathrm{cut}} = 0.05$ in ice. 
    The stochastic interactions 
    are stated as Bremsstrahlung (Brems), photonuclear interaction (Nuclint), electron pair production (Epair), and ionization (Ioniz).
    Two simulations 
    are done to check both multiple scattering methods Molière (MSM) 
    and Highland (MSH).  
    The total distribution is presented only for all stochastic processes including MSM. 
    Since multiple scattering describes the deflection along a 
    continuous energy loss, a single deflection occurs analogous to a stochastic 
    deflection.
    Multiple scattering dominates the deflection. Details are presented in 
    Table~\ref{tab:defl_per_int}.}
    \label{fig:defl_per_int}
\end{figure*}

First, the stochastic deflections described by Van Ginneken~\cite{Van_Ginneken} 
and implemented 
in PROPOSAL are investigated in combination with the two multiple scattering methods. 
For this purpose, $\num{1000}$ muons are propagated from $E_{\text{i}} = \SI{1}{\peta\electronvolt}$ to $E_{\text{f,\,min}} = \SI{1}{\tera\electronvolt}$.
The deflections per interaction are presented 
for each interaction type and the sum over all types in Figure~\ref{fig:defl_per_int}. 
The size of individual deflections 
extend over several orders of magnitude with a median of $\num{3.9e-6}\,\text{deg}$
and a $\SI{95}{\percent}$ central interval of $[\num{2.2e-7}\,\text{deg}, \,\num{1.3e-3}\,\text{deg}]$. 
It follows that the deflections are primarily dominated by multiple scattering, except for a few outliers caused by bremsstrahlung, which 
allows very large energy losses and thus the largest deflections. 
The largest median deflection with the highest $\SI{95}{\percent}$ interval results due to photonuclear interaction.
The median propagation distance with the lower and upper $\SI{95}{\percent}$ 
central interval results to $16.4_{-7.3}^{+24.6}\,\si{\kilo\meter}$.
Detailed values for each interaction type can be found in Table~\ref{tab:defl_per_int}.

\begin{table*}
    \centering 
    \caption{The medians of deflections $\theta$ per interaction from Figure~\ref{fig:defl_per_int} are presented for each stochastic interaction type, the two multiple scattering methods, and the total distribution including 
     MSM with the upper and lower limits of the $\SI{95}{\percent}$ 
    central intervals. The largest median deflection is caused by photonuclear interaction.}
    \begin{tabular}{ccccccc}
        \toprule 
        Brems & Nuclint & Epair & Ioniz & MSM & MSH & Total  \vspace{6pt} \\
        $\theta\,/\,\num{e-5}\,\text{deg}$ & $\theta\,/\,\num{e-5}\,\text{deg}$ & $\theta\,/\,\num{e-5}\,\text{deg}$ & $\theta\,/\,\num{e-5}\,\text{deg}$ & $\theta\,/\,\num{e-5}\,\text{deg}$ & $\theta\,/\,\num{e-5}\,\text{deg}$ & $\theta\,/\,\num{e-5}\,\text{deg}$\\
        \midrule 
        $3.8_{-0.1}^{+297}$ & $11.7_{-4.2}^{+963}$ & $0.1_{-0.02}^{+4.2}$ & $4.4_{-0.1}^{+181}$& $1.2_{-0.05}^{+222}$ & $1.2_{-0.05}^{+225}$ & $0.4_{-0.02}^{+129}$\\ 
        \bottomrule
    \end{tabular}
    \label{tab:defl_per_int}
\end{table*}

\section{Accumulated Muon Deflection}\label{sec:accum_defl}

As shown in Section~\ref{sec:defl_per_int}, the deflection per interaction 
is lower than $\sim\SI{1}{\degree}$ in general. Since these deflections accumulate along the 
propagation path, the angle between the incoming and the outgoing 
muon direction is analyzed. This angle limits the angular resolution 
for neutrino source searches utilizing incoming muons, since there is no information 
about the muon before the detector entry.

\subsection{Comparison with MUSIC and \sc{Geant4}}
First, the deflections in PROPOSAL are compared to 
the tools MUSIC and \textsc{Geant4}.
MUSIC is a tool to simulate the propagation of muons 
through media like rock and water considering the same energy losses as in 
PROPOSAL. Also, the losses are divided into continuous and stochastic 
energy losses by a relative energy cut. Several cross sections, multiple scattering 
methods, and parametrizations for stochastic deflections are 
available. For these studies, the same cross section parametrizations 
as in PROPOSAL are chosen, except those for 
photonuclear interaction \cite{nulcint_bugaev_Shlepin, bugaev_1980_defl,bugaev_1981_defl}. The stochastic deflections are also parametrized by 
Van Ginneken~\cite{Van_Ginneken}. 
The Gaussian 
approximation \cite{HIGHLAND_1975} is set as multiple scattering. 
\textsc{Geant4} is another common toolkit to simulate the passage of particles through 
matter using the same cross section parametrizations except for photonuclear interaction~\cite{Borog:1975_inelastic}. The simulation is very precise and especially 
made for simulations in particle detectors \cite{GEANT4_standard, GEANT4}. 

\begin{figure*}
    \centering 
    \includegraphics{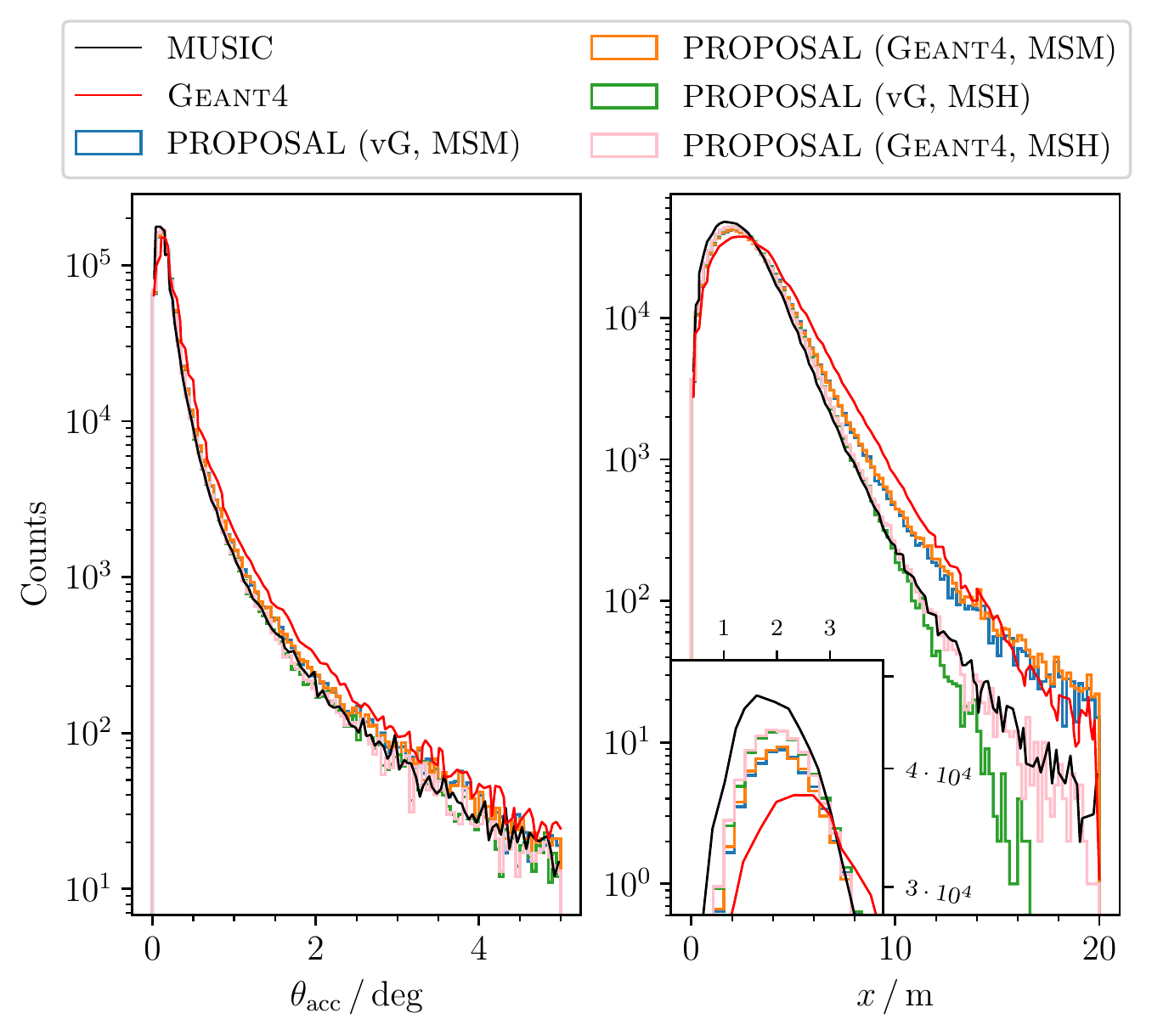}
    \caption{A comparison of the results of MUSIC, \textsc{Geant4}, and PROPOSAL is presented for $\num{e6}$ muons propagated with 
    $E_{\text{i}} = \SI{2}{\tera\electronvolt}$ over a distance of 
    $\SI{3}{\kilo\meter}$ in water. A $v_{\mathrm{cut}} = \num{e-3}$ is set. In PROPOSAL, 
    bremsstrahlung and photonuclear interaction deflections are parametrized either by 
    Van Ginneken (vG) or as in \textsc{Geant4}. Both multiple scattering by Molière (MSM) and Highland (MSH) are checked. 
    Left: The accumulated deflection $\theta_{\mathrm{acc}}$ in degree is very similar in all cases.
    Right: The lateral displacement $x$ in meter depends 
    on the scattering method. MSM leads to larger distances.
    In the zoomed-in figure, the region around the mode of the distributions is presented. 
    The mode of \textsc{Geant4} is shifted to larger deflections.
    Detailed information are given in 
    Table~\ref{tab:compare_MUSIC}. The results for MUSIC and \textsc{Geant4} are taken from 
    Ref.~\cite{comparison_MUSIC_GEANT4_2009}.}
    \label{fig:compare_MUSIC}
\end{figure*}

A comparison of all three tools is shown in Figure~\ref{fig:compare_MUSIC} 
for the 
accumulated deflection angle $\theta_{\text{acc}}$ and the lateral displacement
$x$. 
The propagation of $\num{1000000}$ muons with an 
initial energy of $E_{\mathrm{i}} = \SI{2}{\tera\electronvolt}$ is simulated 
in water with a maximum propagation distance of $\SI{3}{\kilo\meter}$.
Four different settings are studied in PROPOSAL to compare the results with 
the two multiple scattering methods and the different stochastic deflection parametrizations.
The deflection angles are 
similar in all cases. The 
largest displacements are exhibited by \textsc{Geant4} and PROPOSAL with Molière scattering, which 
leads to the largest deflections and thus to a larger displacement. 
PROPOSAL with Highland scattering and MUSIC have less outliers, since large 
deflections are neglected in the Gaussian approximation \cite{HIGHLAND_1975}. 
The combination of Highland and 
Van Ginneken's photonuclear interaction parametrization leads to the smallest 
displacement. This is due to the fact that the angle is sampled from the root mean squared angle in the exponential distribution
in the parameterization for photonuclear interaction by Van Ginneken, which neglects 
outliers to larger angles. 
In general, the lateral displacements differ, although the angles are very similar in all simulations. 
This can be explained by the location of the deflection. If larger deflections occur sooner, 
they lead to further displacements during propagation, although the angle remains the same.

Detailed information are given in Table~\ref{tab:compare_MUSIC}. 
The largest average deflections are obtained in \textsc{Geant4} with 
$\overline{\theta} = \SI{0.27}{\degree}$ 
and $\overline{x} = \SI{3.3}{\meter}$, while MUSIC provides the lowest 
ones with $\overline{\theta} = \SI{0.22}{\degree}$ and 
$\overline{x} = \SI{2.6}{\meter}$.
The results of PROPOSAL lie between 
these two tools. Hence, the mean 
values of all tools are very close to each other and therefore consistent.

\begin{table*}
    \small
    \centering
    \caption{The survival probability $p_{\text{s}}$ defined by the ratio of all 
    muons that reach the propagation distance of $\SI{3}{\kilo\meter}$ and the 
    amount of muons stopping before due to large energy losses and muon decays,
    the mean survived muon 
    energy $\overline{E}_{\text{f}}$, the mean scattered angle $\overline{\theta}$,
    and the mean displacement $\overline{x}$ are presented for all cases from 
    Figure~\ref{fig:compare_MUSIC}. For all means, the standard deviation is given.
    The largest deflection and displacement is observed in the tool \textsc{Geant4}, which has the lowest mean survived energy. The lower the energy, the larger the deflection.}
    \begin{tabular}{l|cc|cccc}
        \toprule
        & MUSIC & \textsc{Geant4} & \multicolumn{4}{c}{PROPOSAL} \\
        &  & & \multicolumn{2}{c}{MSM} & 
        \multicolumn{2}{c}{MSH} \\
        &  &  & vG & \textsc{Geant4} & vG & \textsc{Geant4} \\
        \midrule
        $p_{\text{s}}\,/\,\si{\percent}$ & 77.9 & 79.3 &  \multicolumn{4}{c}{77.9}\\
        $\overline{E}_{\text{f}}\,/\,\si{\giga\electronvolt}$ & 323 & 317 & \multicolumn{4}{c}{331$\pm$178} \\
        $\overline{\theta}\,/\,\si{\degree}$ & 0.22 & 0.27 & 0.24$\pm$0.45 & 0.24$\pm$0.45 & 0.22$\pm$0.35 & 0.22$\pm$0.35   \\
        $\overline{x}\,/\,\si{\meter}$ & 2.6 & 3.3 & 2.9$\pm$2.6 & 2.9$\pm$2.6 & 2.7$\pm$1.6 & 2.7$\pm$1.7  \\
     \bottomrule
    \end{tabular}
    \label{tab:compare_MUSIC}
\end{table*}

\subsection{Data--MC Agreements}
\begin{figure*}
    \centering 
    % \resizebox{0.48\textwidth}{!}{%
    \includegraphics{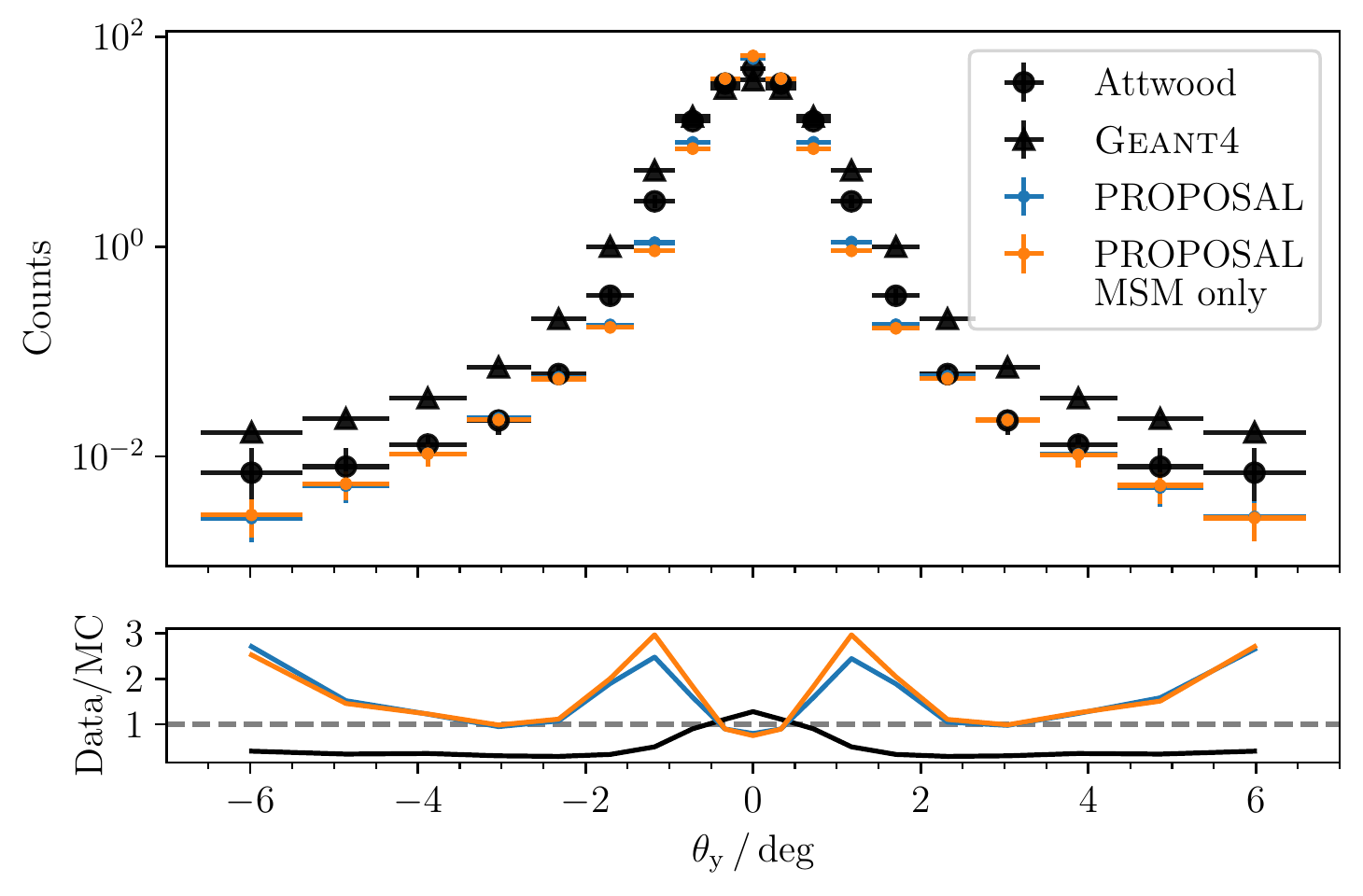}
    % }
    \caption{
    Muons are 
    propagated with $E_{\mathrm{i}} = \SI{199}{\mega\electronvolt}$ through 
    $\SI{109}{\milli\meter}$ of liquid $\text{H}_2$.
    Measured data of Attwood et al. and simulation data of \textsc{Geant4} are taken from Ref.~\cite{attwood_2006}.    
    The figure presents 
    the normalized counts in dependence of the projected scattering angle $\theta_{\mathrm{y}}$ in degree.
    In PROPOSAL, $100$ simulations each with $\num{e5}$ muons are performed for two different settings using the energy cut 
    $v_{\mathrm{cut}} = \num{e-5}$. The blue points present the mean of the simulations considering
    stochastic deflections and Molière scattering (MSM), the orange 
    points
    present the mean of the simulations taking into account only Molière scattering.   
    The uncertainties on the $x$--axis result due to the measured bin widths. The $y$--uncertainties are the standard deviations.   
    The deflections are  
    underestimated in PROPOSAL, except at $\theta_{\mathrm{y}} \approx 
    \SI{0}{\degree}$ and at $\theta_{\mathrm{y}} \approx 
    \SI{3}{\degree}$. At deflections $\SI{2}{\degree} < \theta_{\mathrm{y}} < \SI{5}{\degree}$, 
    the result seems to be more accurate than \textsc{Geant4}'s. The consideration of the stochastic deflections shows no significant influence.}
    \label{fig:attwood_comparison}
\end{figure*}

In the following, two comparisons are performed with measured data for 
different energies and media.
A measurement of muon deflections in low-$Z$ materials was done by Attwood et al. \cite{attwood_2006}. 
From this it can be seen that for $Z < 4$ the scattering angle is overestimated 
by Molière scattering in \textsc{Geant4}. Hence, the lower scattering in PROPOSAL leads 
to a better agreement especially in the region of outliers. The comparison is 
done in liquid $\text{H}_2$ with a thickness of $\SI{109}{\milli\meter}$ and an 
initial particle energy of $E_{\mathrm{i}} = \SI{199}{\mega\electronvolt}$. 
This energy is obtained via the energy-momentum relation of 
a beam momentum of $p = \SI[per-mode=symbol]{168.9}{\mega\electronvolt\per\clight}$
used in Ref.~\cite{attwood_2006}. 
In PROPOSAL, the simulations are done with two different energy cuts $v_{\mathrm{cut}} = \num{e-3}$ and $v_{\mathrm{cut}} = \num{e-5}$, 
but there is no significant difference between the resulting deflections.
Even though in the logarithmic figure the simulation data agree well 
with the measured data, it is clear from the data--MC ratio that the deviations 
are up to $\SI{200}{\percent}$ in some cases. Thus, the deflections are described 
correctly only in a first approximation.
The comparison is presented in Figure~\ref{fig:attwood_comparison}.

\begin{figure*}
    \centering 
    % \resizebox{0.48\textwidth}{!}{%
    \includegraphics{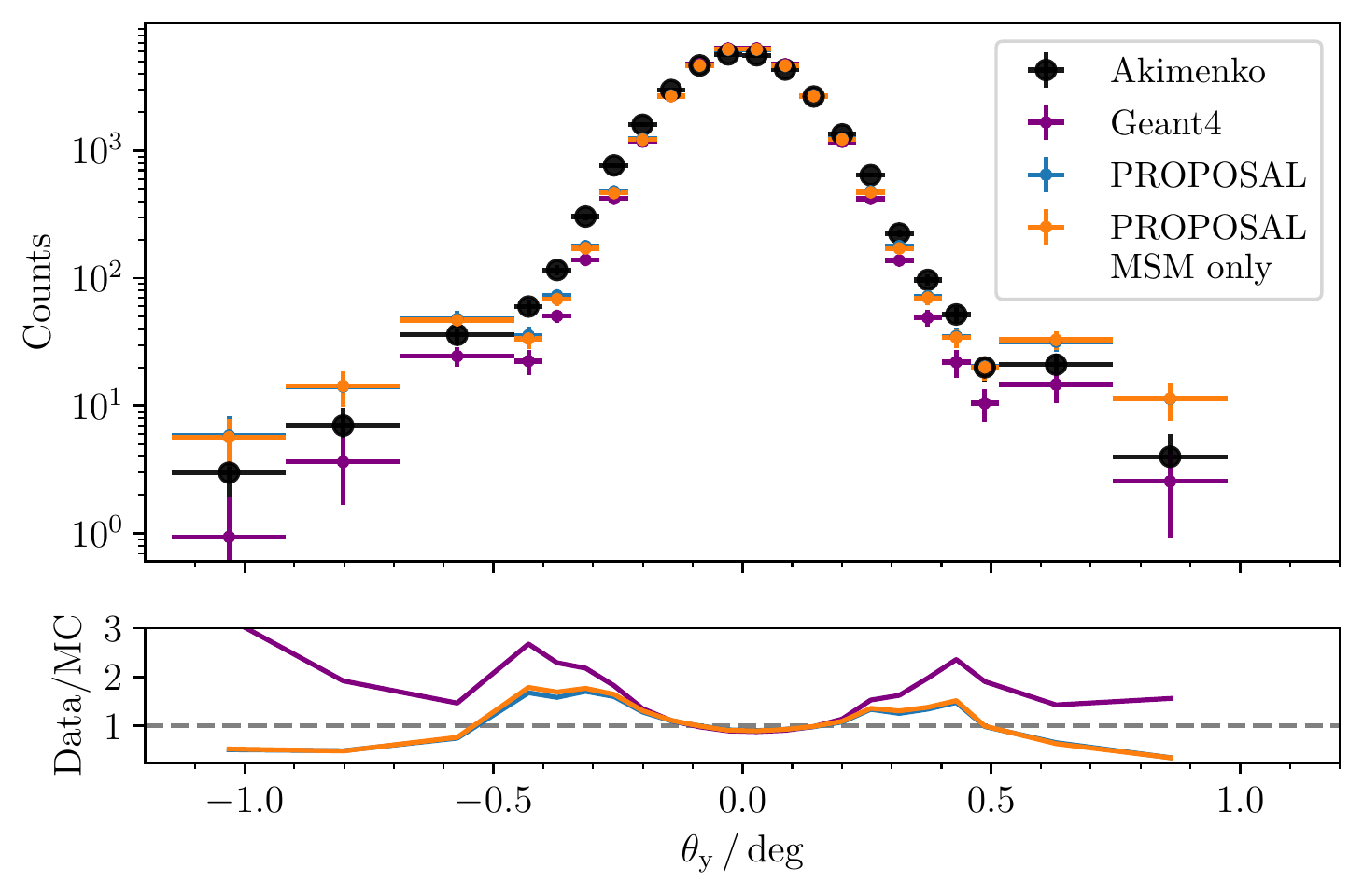}
    % }
    \caption{A comparison between two PROPOSAL simulations
    and measured data by Akimenko et al. \cite{akimenko_1984} 
    is presented for the normalized counts in dependence of the projected scattering angle $\theta_{\mathrm{y}}$ in degree.
    Deflections are simulated with \textsc{Geant4} $\text{v}11.0.3$ as well.
    $\num{100}$ simulations each with $\num{31125}$ muons with 
    $E_{\mathrm{i}} = \SI{7.301}{\giga\electronvolt}$ are propagated through a $\SI{1.44}{\centi\meter}$ 
    copper layer. In PROPOSAL the energy cut 
    $v_{\mathrm{cut}} = \num{e-5}$ is used. The blue points present the mean 
    of the simulation considering stochastic deflections and Molière scattering (MSM), 
    the orange points
    present the mean of the simulation taking into account only 
    Molière scattering. The uncertainties on the $x$--axis result due to the measured bin widths. The $y$--uncertainties are the standard deviations for PROPOSAL and the Poisson errors with $\sqrt{N}$ for Akimenko, with $N$ as number of counts. More large deflections are simulated by PROPOSAL.
    Considering stochastic deflections has no significant impact.}
    \label{fig:akimenko_comparison}
\end{figure*}

The second measurement of muon deflections is done for higher energetic muons 
of $p = \SI[per-mode=symbol]{7.3}{\giga\electronvolt\per\clight}$ by Akimenko et al. \cite{akimenko_1984}.
In total, $\num{31125}$ muons are propagated through a $\SI{1.44}{\centi\meter}$ thick 
copper layer. Again, the two energy cuts mentioned before and the effect of stochastic deflections 
in comparison with Molière scattering only are checked. 
Neither between the two energy cuts, nor when using the stochastic deflection a significant difference occurs. 
PROPOSAL simulates more large deflections than observed in these data. This observation differs from the comparison with Attwood, 
in which less higher deflections are simulated. In general, the higher muon energy 
leads to smaller deflections. 
Also simulations with \textsc{Geant4} $\text{v}11.0.3$ using the 
default settings and the \textit{PhysicsLists} \texttt{QBBC} and \texttt{FTFP\_BERT} 
are performed. At angles larger than $\SI{0.2}{\degree}$, 
\textsc{Geant4} simulates less large deflections than PROPOSAL and 
less than expected in the data. Similar to the comparison with Attwood, 
data--MC mismatches larger than $\SI{100}{\percent}$ are observed. 
There are no differences between the two \textit{PhysicsLists} in the resulting 
deflections.  
The result is presented in Figure~\ref{fig:akimenko_comparison}.

\subsection{Muon Deflection Impact on Angular Resolutions}
\begin{figure*}
    \centering 
    \includegraphics{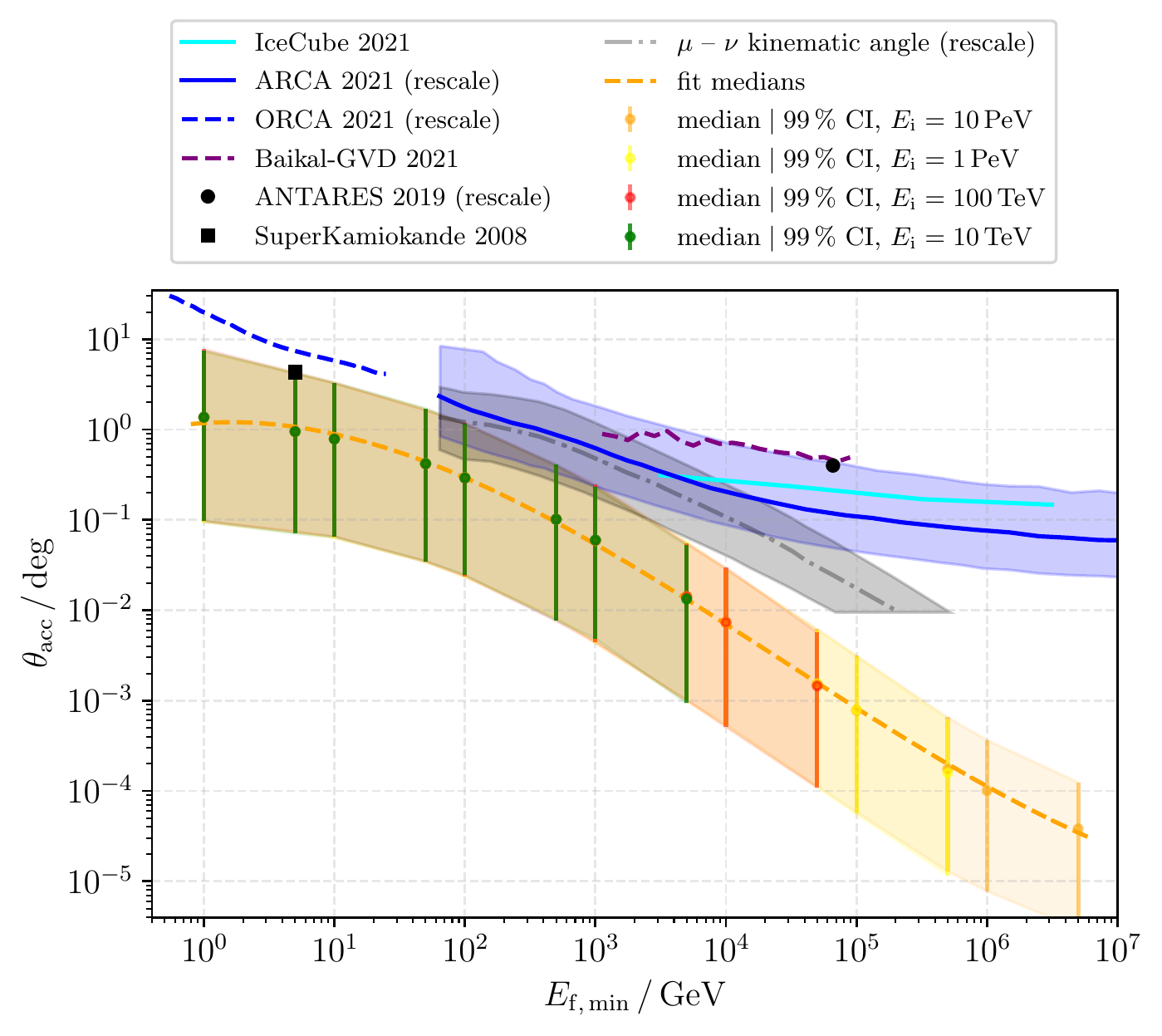}
    \caption{The median of the accumulated deflection $\theta_{\text{acc}}$ in degree 
    with a $\SI{99}{\percent}$ 
    central interval is shown for four different initial energies $E_{\text{i}}$. 
    Each data set includes more than $\num{50000}$ events with the requirement 
    that the true final muon energy $E_{\text{f}}$ is at most 
    $\SI{10}{\percent}$ below the set final energy $E_{\text{f,\,min}}$,   
    $E_{\text{f}} > E_{\text{f,\,min}} \cdot 0.9$. The energy cuts are $e_{\mathrm{cut}} = \SI{500}{\mega\electronvolt}$ and $v_{\mathrm{cut}} = 0.05$, and 
    Molière scattering is chosen. Simulations are performed in ice, the deviations 
    of the medians in a water-based simulation are smaller than $\SI{1}{\percent}$.
    Since the medians overlap for different initial energies, there is no 
    strong impact of the initial energy on the median deflection. These 
    medians can be fit by a third degree polynomial in the log-space as 
    shown in Eq.~\eqref{eqn:fit_median}. The kinematic angle between the muon and 
    neutrino is taken from Ref.~\cite{KM3NeT_Resolution2021}. Since the kinematic angle and 
    the angular resolution of ARCA taken from Ref.~\cite{KM3NeT_Resolution2021} are 
    presented in dependence of the neutrino energy as well as the resolutions of ORCA~\cite{ORCA_Resolution2021} 
    and ANTARES~\cite{ANTARES_Resolution2019}, a rescaling to the muon energy is performed 
    using the average energy transfer to the nucleus \cite{GANDHI199681}. 
    For energies 
    $E_{\text{f}} \leq \SI{1}{\tera\electronvolt}$, there is a minimal influence of the deflection on the angular resolution of 
    ARCA and at $E_{\mathrm{f}} = \SI{5}{\giga\electronvolt}$ the upper limit of the deflections affects the resolution 
    of SuperKamiokande~\cite{SuperKamiokande_Resolution2008}. The resolutions shown by IceCube~\cite{IceCube_Resolution2021}, ORCA, 
    Baikal~\cite{Baikal_Resolution2021} and ANTARES are not impacted, but no uncertainty bands are given for these either.
    The exact values and the propagated distances are presented in Table~\ref{tab:final_values}.}
    \label{fig:fit_median}
\end{figure*}

For neutrino source searches based on muons entering the detector, 
it is important to study the impact of the muon 
deflection on the angular resolution to estimate whether or not this needs to be 
taken into account as an additional source of uncertainty.
For this purpose, four different initial energies 
from $E_{\text{i}} = \SI{10}{\tera\electronvolt}$ to 
$E_{\text{i}} = \SI{10}{\peta\electronvolt}$ are used and the final 
energy is set to $E_{\text{f,\,min}} \geq \SI{1}{\giga\electronvolt}$ with 
$E_{\text{f,\,min}} < E_{\text{i}}$ for each simulation. 
This energy range covers the muon energies typically measured in 
neutrino experiments.
In total $\num{44}$ simulations are performed. 
To compare the results of these simulations, the medians of the deflection distributions 
with a $\SI{99}{\percent}$ central interval are presented in 
Figure~\ref{fig:fit_median}.
The lower the final muon energy, the larger the accumulated deflection. 
For energies $E_{\text{f}} = \SI{1}{\peta\electronvolt}$, the median deflection 
is $\num{e-4}\,\text{deg}$. For energies $E_{\text{f}} = \SI{100}{\giga\electronvolt}$, 
angles larger than $\SI{1}{\degree}$ are possible. 
For energies  
$E_{\text{f}} \leq \SI{1}{\tera\electronvolt}$, 
there is a small overlap of the deflection with the angular resolution of KM3NeT/ARCA 
\cite{KM3NeT_Resolution2021, KM3NeT_Resolution2016}. 
At low energies of $E_{\mathrm{f}} = \SI{5}{\giga\electronvolt}$, the upper limit of the 
deflections affects the resolution of SuperKamiokande~\cite{SuperKamiokande_Resolution2008}. 
The kinematic scattering angle between the incident neutrino and the produced muon is 
larger than the deflection in the presented region from $\SI{60}{\giga\electronvolt}$
to $\SI{200}{\tera\electronvolt}$. For energies below $\SI{2}{\tera\electronvolt}$,  
the muon deflections are of the same order of magnitude as the kinematic angle and thus become increasingly relevant.
Here it must be noted that the kinematic angle 
and the resolution of ARCA in Ref.~\cite{KM3NeT_Resolution2021} as well as 
the resolutions of ORCA (part of KM3NeT optimized to study atmospheric neutrinos in the $\si{\giga\electronvolt}$ energy range) \cite{ORCA_Resolution2021} and ANTARES~\cite{ANTARES_Resolution2019} are presented in dependence of the neutrino energy. 
Hence, a rescaling to the muon energy is applied using the average energy transfer of the neutrino to 
the nucleus \cite{GANDHI199681}. This shifts the curves to lower energies. 
Since all of these simulations are done 
in ice, the same simulations are done in water to compare the results for 
water-based experiments. The deviations of the medians
are less than $\SI{1}{\percent}$ for all energies and therefore not shown.
The accumulated deflections and also the propagated distances 
of Figure~\ref{fig:fit_median} are presented in Table~\ref{tab:final_values}.
Muons are able to propagate various distances for a fixed final muon energy
depending on the stochasticity of the energy losses.

Note that the distribution of deflection angles at a given final energy $E_{\mathrm{f}}$ in Figure~\ref{fig:fit_median} overlap for differing initial energies. This result indicates that the total deflection of a muon 
primarily depends on the final muon energy.
The initial muon energy is nearly irrelevant. 
Hence, the reconstructed muon 
energy in a detector can be used to estimate the deflection. For this 
purpose, a polynomial of degree three as 
\begin{equation}
     f(x) = a \cdot x^3 + b \cdot x^2 + c \cdot x + d \,,
    \label{eqn:fit_median}
\end{equation} 
can be used with the parameters 

\begin{align*}
        a =& +0.0176\pm 0.0018\,,  & c =& +0.0929\pm 0.0527\,,\\
        b =& -0.2328\pm 0.0185\,,  & d =& +0.0726\pm 0.0404\,,
\end{align*}

in the logarithmic space via 
\begin{align}
    g(x) =& 10^{f(x)}\,, & x =& \log_{10}\left(\frac{E_{\text{f}}}{\si{\giga\electronvolt}}\right)\,.
\end{align}
In general, the function $f(x)$ in Eq.~\eqref{eqn:fit_median} describes the median 
deflection of a muon after a propagated distance in ice for a given, respectively measured energy 
to estimate the deflection before the detector entry. 
This equation is valid for muon energies between 
$\SI{1}{\giga\electronvolt}$ and $\SI{50}{\peta\electronvolt}$. 
Basically, it should be mentioned here that the data--MC comparisons shown 
earlier are for energies of $E_{\mathrm{i}} = \SI{199}{\mega\electronvolt}$ and 
$E_{\mathrm{i}} = \SI{7.3}{\giga\electronvolt}$, which are much lower than the 
energies in these simulations.

\begin{figure}
    \resizebox{0.48\textwidth}{!}{%
    \includegraphics{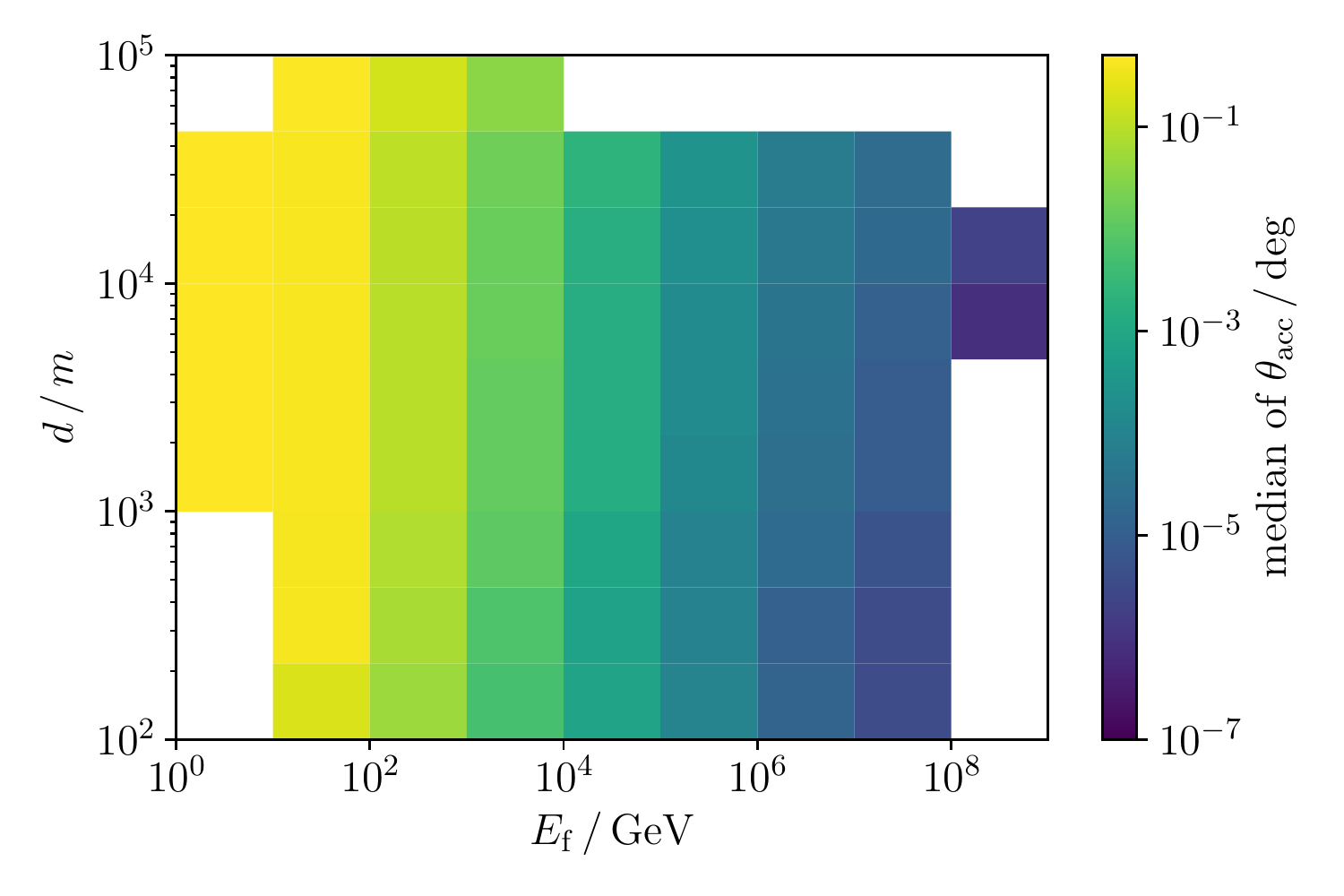}
    }
    \caption{The median of the accumulated deflection $\theta_{\mathrm{acc}}$ is presented in degree as 
    a function of the final muon energy $E_{\mathrm{f}}$ and the propagation distance $d$. 
    The result is presented 
    for $\num{50370}$ muons for a propagation through ice. An energy cut of $v_{\mathrm{cut}} = \num{1e-3}$, MSM and the deflection 
    parametrizations of vG are used. The initial energies are not fixed. The muon deflection 
    depends primarily only on the final muon energy. Even the propagation distance 
    is negligible if the initial muon energy is unknown.}
    \label{fig:3D}
\end{figure}

To analyze the impact of the propagation distance on the muon deflection, 
another simulation is done. This time, the initial energies are not fixed and 
sampled from an atmospheric muon flux at sea level by \cite{gaisser1990} 
with a weighting of $E^{3.7}$ for 
energies from $\SI{10}{\giga\electronvolt}$ to $\SI{1e10}{\giga\electronvolt}$.
The final energies are sampled similar. The resulting deflections are presented in 
Figure~\ref{fig:3D}. From this follows, that the median deflection 
is not impacted by the propagation distance, if the initial muon energy is 
unknown. This is a realistic scenario for example for a neutrino telescope, since 
the only known value is the reconstructed muon energy at the detector entry. 
There are no information about the initial energy and the propagation distance. 
Finally, the muon deflection can be estimated only by the reconstructed muon 
energy.    

\subsection{Relevance for Muography}
Muography is a technique to study the inside of structures with a 
wide field of applications such as the monitoring of volcanic activity and many 
more. For this, the atmospheric muon flux is measured with a detector located 
below or even behind the object, which is visualized. 
The muon flux count rates then depend on the densities of the materials, 
the higher the density, the stronger the attenuation. 
Based on this, conclusions can be drawn about materials and cavities in the 
respective object. Sufficient statistics in reasonable time are obtained for 
muons in the $\si{\giga\electronvolt}$ energy range \cite{muography_2019}. 

Angular resolutions of these detectors are below $\SI{0.6}{\degree}$, which 
is on the order of magnitude of the muon deflection at $\si{\giga\electronvolt}$
muon energies. Multiple scattering of muon deflection in several media is 
also studied in \cite{muography_2019, Geant4_scattering}. The resulting 
deflections are about $\SI{1}{\degree}$, similar to the deflections 
expected with PROPOSAL in water and ice. Hence, the scattering 
of muons can be a limiting factor for the angular resolution in 
muography at energies of a few $\si{\giga\electronvolt}$.

\section{Conclusion}\label{sec:conclusion}

Stochastic deflection, recently implemented in PROPOSAL 7.3.0, is 
used to study the muon deflection per interaction. The deflection 
is dominated by multiple scattering except for a few stochastic 
outliers by bremsstrahlung. These angles are lower than
$\sim\SI{1}{\degree}$. 

The results of PROPOSAL are compared with the common tools MUSIC and 
\textsc{Geant4} and they are in good agreement. In low-$Z$ materials, the region of outlier 
deflections fits the measured data better with PROPOSAL, than \textsc{Geant4}. 
A second data comparison points out that PROPOSAL simulates more 
large deflections at higher muon energies and less at lower 
energies. 
In the data--MC ratio deviations up 
to a factor of $3$ are observed. 
This points out that still improvements are required in the deflection 
parameterizations and 
in the multiple scattering, respectively.
Since the presented measurements of the muon deflection are based on muon energies 
lower $\SI{10}{\giga\electronvolt}$, deflection measurements of muons with energies 
up to $\si{\tera\electronvolt}$ or even higher are required to validate the 
results at higher energies.

The median accumulated deflection depends primarily on the final muon energy, which can be interpreted as the muon energy at detector entry 
in neutrino telescopes or other muon detectors.
The outcome is fit by a polynomial and can be used for 
a theoretical estimation of the muon deflection in water and ice.
Since the result can be interpreted as the deflection before the detector entry, it defines a lower limit on the directional resolution.
At energies lower $\SI{1}{\tera\electronvolt}$, there is potentially a small impact of the muon deflection on the angular 
resolution of KM3NeT.

\begin{acknowledgement}
  This work has been supported by the DFG, Collaborative Research Center SFB $876$
  under the project C3 
  (https://sfb876.tu-dortmund.de) and the SFB $1491$ (https://www.sfb1491.ruhr-uni-bochum.de).
  We also acknowledge the funding by the DFG under the project number SA~$3867/$2-1.
\end{acknowledgement}

\begin{table*}
    \centering
    \caption{The median values for the accumulated deflection $\theta_{\mathrm{acc}}$ in degree and the propagated distances 
    $d$ in meter are presented for Figure~\ref{fig:fit_median}. The upper and lower values indicate the upper und lower 
    $\SI{99}{\percent}$ intervals around the median. It turns out that the median deflection depends primarily 
    only on the final muon energy $E_{\mathrm{f,\,min}}$.}
    \label{tab:final_values}       % Give a unique label
    \begin{tabular}{r|ll|ll|ll|ll}
    \hline\noalign{\smallskip}
    & \multicolumn{2}{c|}{$E_{\mathrm{i}} = \SI{10}{\peta\electronvolt}$} & \multicolumn{2}{c|}{$E_{\mathrm{i}} = \SI{1}{\peta\electronvolt}$} 
    & \multicolumn{2}{c|}{$E_{\mathrm{i}} = \SI{100}{\tera\electronvolt}$} & \multicolumn{2}{c}{$E_{\mathrm{i}} = \SI{10}{\tera\electronvolt}$} \\
    $E_{\mathrm{f,\,min}}\,/\,\si{\giga\electronvolt}$ & $\theta_{\text{acc}}\,/\,\si{\degree}$ & $d\,/\,\si{\kilo\meter}$  
    & $\theta_{\text{acc}}\,/\,\si{\degree}$ & $d\,/\,\si{\kilo\meter}$ 
    & $\theta_{\text{acc}}\,/\,\si{\degree}$ & $d\,/\,\si{\kilo\meter}$
    & $\theta_{\text{acc}}\,/\,\si{\degree}$ & $d\,/\,\si{\kilo\meter}$\\
    \noalign{\smallskip}\hline\noalign{\smallskip}
    \rule{0pt}{2.6ex}\num{1}        & $1.36_{0.10}^{7.33}$ & $23.7_{9.10}^{37.8}$ & $1.36_{0.09}^{7.38}$ & $18.7_{6.05}^{30.4}$ & $1.37_{0.10}^{7.75}$ & $13.5_{3.60}^{21.9}$ & $1.37_{0.10}^{7.47}$ & $7.91_{1.65}^{12.2}$ \\
    \rule{0pt}{2.6ex}\num{5}        & $9.48_{0.74}^{41.9}\times 10^{-1}$ & $23.8_{8.99}^{37.6}$ & $9.53_{0.74}^{42.3}\times 10^{-1}$ & $18.7_{5.97}^{30.3}$ & $9.52_{0.74}^{42.7}\times 10^{-1}$ & $13.4_{3.48}^{21.9}$ & $9.50_{0.71}^{42.3}\times 10^{-1}$ & $7.89_{1.63}^{12.2}$ \\
    \rule{0pt}{2.6ex}\num{10}       & $7.88_{0.64}^{33.1}\times 10^{-1}$ & $23.8_{9.15}^{37.8}$ & $7.93_{0.63}^{32.0}\times 10^{-1}$ & $18.7_{6.19}^{30.1}$ & $7.89_{0.65}^{33.0}\times 10^{-1}$ & $13.4_{3.62}^{21.9}$ & $7.85_{0.65}^{33.0}\times 10^{-1}$ & $7.84_{1.58}^{12.1}$ \\
    \rule{0pt}{2.6ex}\num{50}       & $4.19_{0.34}^{16.5}\times 10^{-1}$ & $23.7_{9.10}^{37.6}$ & $4.21_{0.34}^{16.8}\times 10^{-1}$ & $18.5_{6.08}^{30.0}$ & $4.18_{0.35}^{16.5}\times 10^{-1}$ & $13.2_{3.58}^{21.7}$ & $4.20_{0.35}^{17.2}\times 10^{-1}$ & $7.72_{1.48}^{12.0}$ \\
    \rule{0pt}{2.6ex}\num{100}      & $2.95_{0.24}^{11.7}\times 10^{-1}$ & $23.5_{9.00}^{37.2}$ & $2.92_{0.23}^{11.0}\times 10^{-1}$ & $18.4_{6.02}^{29.9}$ & $2.91_{0.24}^{11.8}\times 10^{-1}$ & $13.1_{3.54}^{21.6}$ & $2.92_{2.44}^{11.5}\times 10^{-1}$ & $7.55_{1.46}^{11.8}$ \\
    \rule{0pt}{2.6ex}\num{500}      & $1.01_{0.08}^{3.92}\times 10^{-1}$ & $22.6_{8.33}^{36.3}$ & $1.02_{0.08}^{3.86}\times 10^{-1}$ & $17.5_{5.44}^{28.9}$ & $1.02_{0.07}^{4.06}\times 10^{-1}$ & $12.2_{2.87}^{20.5}$ & $1.02_{0.08}^{4.09}\times 10^{-1}$ & $6.68_{1.01}^{10.7}$ \\
    \rule{0pt}{2.6ex}\num{1000}     & $5.90_{0.44}^{22.7}\times 10^{-2}$ & $21.8_{7.65}^{35.4}$ & $5.92_{0.44}^{2.29}\times 10^{-2}$ & $16.8_{4.85}^{27.9}$ & $5.92_{0.44}^{24.7}\times 10^{-2}$ & $11.5_{2.38}^{19.4}$ & $6.00_{0.48}^{23.1}\times 10^{-2}$ & $5.86_{0.78}^{9.54}$ \\
    \rule{0pt}{2.6ex}\num{5000}     & $1.41_{0.10}^{5.42}\times 10^{-2}$ & $19.0_{6.16}^{31.6}$ & $1.42_{0.10}^{5.78}\times 10^{-2}$ & $13.9_{3.38}^{24.0}$ & $1.42_{0.10}^{5.54}\times 10^{-2}$ & $8.58_{1.20}^{15.2}$ & $1.33_{0.09}^{5.30}\times 10^{-2}$ & $2.54_{0.17}^{4.28}$ \\
    \rule{0pt}{2.6ex}\num{10000}    & $7.33_{0.55}^{28.3}\times 10^{-3}$ & $17.4_{5.11}^{29.4}$ & $7.36_{0.52}^{28.8}\times 10^{-3}$ & $12.4_{2.60}^{21.8}$ & $7.37_{0.51}^{29.5}\times 10^{-3}$ & $7.03_{0.76}^{12.7}$ & --- & --- \\
    \rule{0pt}{2.6ex}\num{50000}    & $1.53_{0.11}^{6.23}\times 10^{-3}$ & $13.6_{3.25}^{24.0}$ & $1.53_{0.11}^{6.01}\times 10^{-3}$ & $8.56_{1.16}^{15.7}$ & $1.46_{0.11}^{5.70}\times 10^{-3}$ & $2.75_{0.17}^{5.02}$ & --- & --- \\
    \rule{0pt}{2.6ex}\num{100000}   & $7.80_{0.58}^{31.4}\times 10^{-4}$ & $11.9_{2.49}^{21.4}$ & $7.90_{0.55}^{31.0}\times 10^{-4}$ & $6.89_{0.73}^{12.8}$ & --- & --- & --- & --- \\
    \rule{0pt}{2.6ex}\num{500000}   & $1.75_{0.13}^{6.53}\times 10^{-4}$ & $8.16_{1.11}^{15.3}$ & $1.58_{0.12}^{6.14}\times 10^{-4}$ & $2.67_{0.16}^{5.04}$ & --- & --- & --- & --- \\
    \rule{0pt}{2.6ex}\num{1000000}  & $1.01_{0.08}^{3.65}\times 10^{-4}$ & $6.54_{0.64}^{12.4}$ & --- & --- & --- & --- & --- & --- \\
    \rule{0pt}{2.6ex}\num{50000000} & $3.81_{0.30}^{12.5}\times 10^{-5}$ & $2.55_{0.13}^{4.96}$ & --- & --- & --- & --- & --- & --- \\
    \noalign{\smallskip}\hline
    \end{tabular}
    % Or use
    % \vspace*{5cm}  % with the correct table height
    \end{table*}

\printbibliography

\end{document}